\documentclass[twoside]{article}

\usepackage{PRIMEarxiv}

\usepackage[utf8]{inputenc} 
\usepackage[T1]{fontenc}    
\usepackage{hyperref}       
\usepackage{url}            
\usepackage{booktabs}       
\usepackage{amsfonts}       
\usepackage{nicefrac}       
\usepackage{microtype}      
\usepackage{lipsum}
\usepackage{fancyhdr}       
\usepackage{graphicx}       
\graphicspath{{media/}}     

\usepackage{amsmath}
\usepackage{multirow}

\title{Systematic Evaluation of Wavelet-Based Denoising for MRI Brain Images: Optimal Configurations and Performance Benchmarks
}

\author{
  Asadullah Bin Rahman, Masud Ibn Afjal, Md. Abdulla Al Mamun\\
  Department of Computer Science and Engineering\\
  Hajee Mohammad Danesh Science and Technology University (HSTU)\\
  Dinajpur -- 5200, Bangladesh\\
  \texttt{galib.cse.17020221@std.hstu.ac.bd, {\{masud, mamun\}@hstu.ac.bd}} \\
}

\begin{document}
\maketitle

\begin{abstract}
Medical imaging modalities including magnetic resonance imaging (MRI), computed tomography (CT), and ultrasound are essential for accurate diagnosis and treatment planning in modern healthcare. However, noise contamination during image acquisition and processing frequently degrades image quality, obscuring critical diagnostic details and compromising clinical decision-making. Additionally, enhancement techniques such as histogram equalization may inadvertently amplify existing noise artifacts, including salt-and-pepper distortions. This study investigates wavelet transform-based denoising methods for effective noise mitigation in medical images, with the primary objective of identifying optimal combinations of threshold values, decomposition levels, and wavelet types to achieve superior denoising performance and enhanced diagnostic accuracy. Through systematic evaluation across various noise conditions, the research demonstrates that the \texttt{bior6.8} biorthogonal wavelet with universal thresholding at decomposition levels 2-3 consistently achieves optimal denoising performance, providing significant noise reduction while preserving essential anatomical structures and diagnostic features critical for clinical applications.
\end{abstract}


\section{Introduction}

Noise contamination in medical images presents significant challenges for diagnostic accuracy and clinical decision-making. Multiple sources contribute to this degradation: thermal noise in MRI systems, quantum noise in CT scans due to limited X-ray photon detection, and speckle noise inherent to ultrasound imaging. These artifacts manifest as random fluctuations, blurring, and structural distortions that complicate radiological interpretation.

Traditional denoising approaches, including linear filtering and non-local means methods, have demonstrated limited efficacy with complex noise patterns and often introduce unwanted artifacts or over-smoothing. Recent advances have explored hybrid methodologies: Kollem et al. \cite{kollem_novel_2023} proposed a diffusivity function-based PDE approach incorporating Quaternion Wavelet Transform with generalized cross-validation for threshold selection, effectively preserving edges while reducing noise. Wavelet Neural Networks (WNNs) integrate wavelets as preprocessing steps or activation functions, enabling hybrid denoising approaches \cite{guo_review_2022}. Additionally, combined Wavelet Transform and Singular Value Decomposition methods have shown improved signal-to-noise ratio performance \cite{patil_efficient_2023}. Zhang et al. \cite{zhang_study_2022} introduced a fractional-order total variation model leveraging differential operators and sparrow search algorithms to balance noise removal and texture preservation.

Deep learning techniques, particularly autoencoders and convolutional neural networks, have revolutionized medical image denoising by learning complex noise patterns. Walid et al. \cite{el-shafai_efficient_2022} presented CADTra, a deep-learning-based autoencoder approach utilizing a Gaussian-distribution based loss function to improve noise removal and pneumonia classification. Comprehensive surveys like that of Izadi et al. \cite{izadi_image_2023} highlight the progression of deep denoisers, covering benchmark datasets, evaluation metrics, and both supervised and unsupervised methods. Hybrid deep learning approaches, such as MWDCNN \cite{tian_multi-stage_2023}, combine CNNs, Wavelet Transforms, and residual blocks to achieve superior denoising performance in natural images, with potential applications in medical imaging. However, these methods require extensive training data, significant computational resources, and lack interpretability—limitations that constrain their clinical applicability \cite{gondara_medical_2016}.

This research focuses on wavelet transform techniques for MRI brain image denoising, building upon established foundations while providing systematic evaluation of optimal configurations. The key contributions include:

\begin{itemize}
    \item Identification of optimal wavelet configurations (sym4, db3, and bior6.8) for MRI denoising across various noise levels
    \item Comprehensive performance evaluation using PSNR and SSIM metrics
    \item Establishment of baseline performance benchmarks for future deep learning integration
\end{itemize}

In the following sections, we discuss our approach in detail. Section \ref{methodology} describes the methodology. Section \ref{result} presents an in-depth analysis of the dataset, parameters, and results. Finally, Section \ref{conclusion} highlights conclusions and future research directions.

\section{Methodology} \label{methodology}

\subsection{Wavelet Transform for Image Denoising}

Wavelet decomposition provides several advantages for medical image denoising:

\begin{enumerate}
    \item \textbf{Multi-resolution Analysis}: Simultaneous analysis at different resolution levels preserves important features while removing noise
    \item \textbf{Localization Properties}: Dual frequency-spatial localization enables precise noise identification while maintaining image integrity
    \item \textbf{Sparse Representation}: Medical images exhibit sparse wavelet domain representations, facilitating effective noise-signal separation
    \item \textbf{Computational Efficiency}: Deterministic approach requiring no training data or extensive computation
    \item \textbf{Edge Preservation}: Effective boundary information retention crucial for diagnostic significance
\end{enumerate}

\begin{figure}[!hbt] 
\centering 
\includegraphics[width = 0.6\linewidth, keepaspectratio]{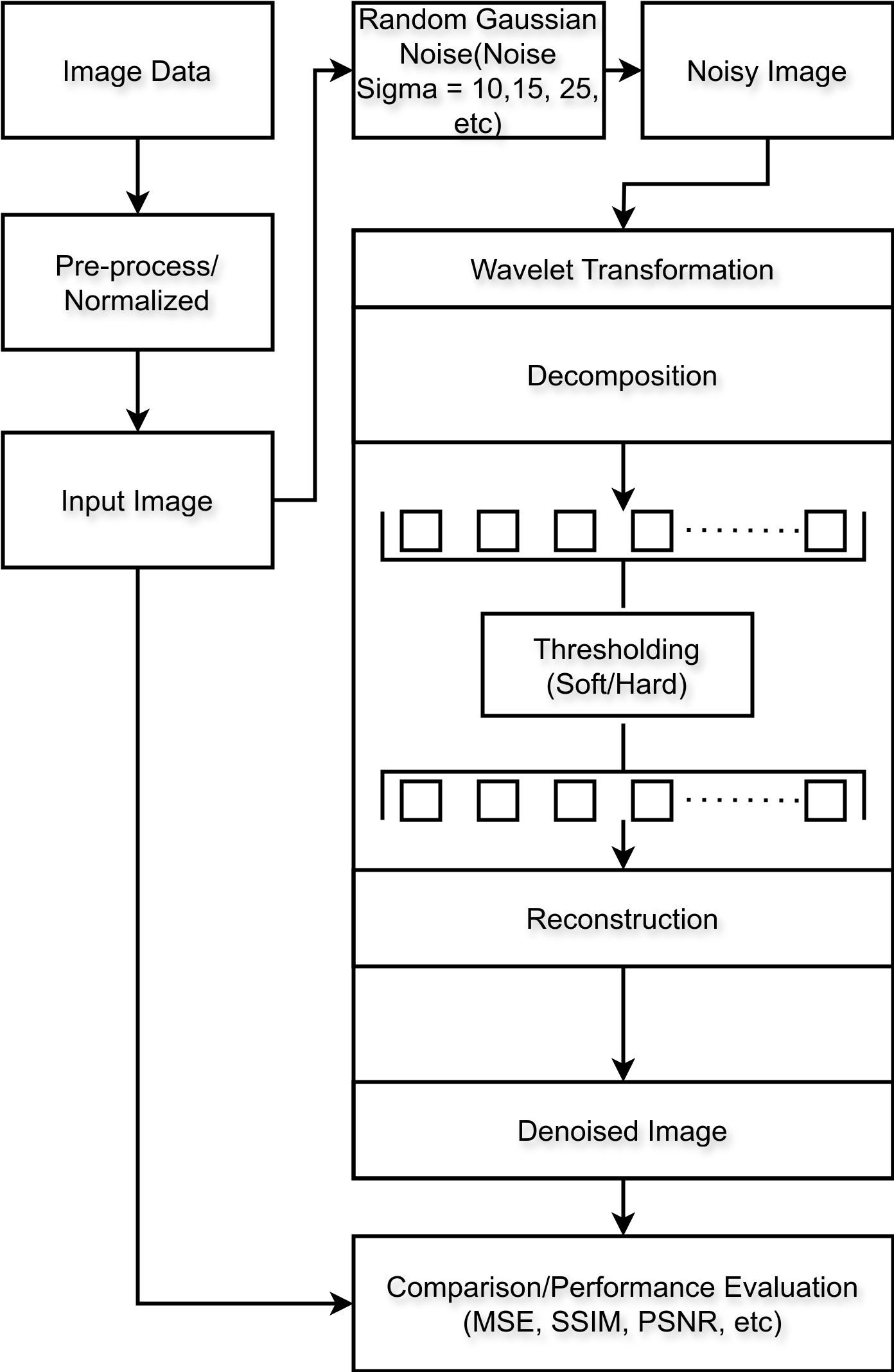} 
\caption{A detailed system flowchart} 
\label{fig:systemFlow} 
\end{figure}

The detailed system workflow is depicted in Figure \ref{fig:systemFlow}.

\subsection{Dataset and Preprocessing}

The study utilized the MRI image dataset sourced from figshare \cite{jun_cheng_brain_2017}, consisting of 3,064 brain tumor MRI images. Preprocessing steps included:

\begin{enumerate}
    \item \textbf{Image Retrieval}: Selected images were loaded into the experimental setup
    \item \textbf{Normalization}: Pixel intensities scaled to 0–255 range for processing consistency
    \item \textbf{Noise Addition}: Gaussian noise introduced with mean ($\mu$) = 0 and standard deviations ($\sigma$) = 10, 15, 25 to simulate realistic noise conditions
\end{enumerate}

\subsection{Wavelet Selection}

Three wavelet families were evaluated based on their suitability for image processing:

\begin{itemize}
    \item \textbf{Daubechies (db3)}: The db3 wavelet belongs to the Daubechies family, characterized by compact support and orthogonality. With three vanishing moments, it effectively captures polynomial behavior in data while balancing detail preservation and noise reduction.

    \item \textbf{Biorthogonal (bior6.8)}: The bior6.8 wavelet offers separate scaling and wavelet functions, with 6 vanishing moments for the scaling function and 8 for the wavelet function. This biorthogonal wavelet provides superior reconstruction properties with symmetry, maintaining sharp edges while reducing noise.

    \item \textbf{Symlet (sym4)}: The sym4 wavelet, a modification of the Daubechies wavelet, features improved symmetry with four vanishing moments. This symmetry provides advantages in image processing tasks, offering balanced noise reduction and edge retention.
\end{itemize}

\subsection{Threshold Calculation Methods}

Calculating the optimal threshold is critical for effectively distinguishing between signal and noise in wavelet-based denoising. Two methods were employed:

\begin{enumerate}
    \item \textbf{Bayes Method}: This approach minimizes the mean squared error (MSE) between the estimated and true images by leveraging the statistical properties of the noise \cite{chipman_adaptive_1997}. The optimal threshold value is given by:
    \begin{equation}
        \tau_{\text{bayes}} = 
        \begin{cases} 
            \frac{\sigma_{\text{noise}}^2}{\sigma_{\text{signal}}} & \text{if } \sigma_{\text{signal}} > 0 \\
            \max(|c_j|) & \text{otherwise}
        \end{cases}
    \end{equation}

    where $\sigma_{\text{signal}} = \sqrt{\max(\text{Var}(c) - \sigma_{\text{noise}}^2, 0)}$ estimates the signal standard deviation from the wavelet coefficients $c$, and $\sigma_{\text{noise}}$ is the noise standard deviation.

    \item \textbf{Universal Method}: The Universal method provides a global threshold, simplifying the denoising process \cite{donoho_ideal_1994}. The threshold is defined as:
    \begin{equation}
        \tau_{\text{universal}} = \sigma_{\text{noise}} \sqrt{2 \log(N)}
    \end{equation}

    where $N$ is the number of wavelet coefficients.
\end{enumerate}

\subsection{Thresholding Techniques}

Thresholding modifies wavelet coefficients to suppress noise while retaining significant signal components:

\begin{enumerate}
    \item \textbf{Hard Thresholding}: Coefficients below the threshold ($\tau$) are set to zero, while those above remain unchanged \cite{donoho_-noising_1995}:
    \begin{equation}
        \hat{c}_j = 
            \begin{cases} 
                c_j & \text{if } |c_j| > \tau \\
                0 & \text{otherwise}
            \end{cases}
    \end{equation}

    \item \textbf{Soft Thresholding}: Coefficients are both shrunk and thresholded \cite{donoho_-noising_1995}:
    \begin{equation}
        \hat{c}_j = 
            \begin{cases} 
                0 & \text{for } |c_j| \leq \tau \\
                c_j - \tau & \text{for } c_j > \tau \\
            \end{cases}
    \end{equation}
\end{enumerate}

where $c_j$ is the original wavelet coefficient, $\tau$ is the optimal threshold value, and $\hat{c}_j$ is the modified coefficient after thresholding.

\subsection{Performance Metrics}

The following metrics were employed for comprehensive evaluation:




\textbf{Peak Signal-to-Noise Ratio (PSNR)} measures maximum signal power relative to noise \cite{wang_mean_2009}:

\begin{equation}
\text{PSNR} = 10 \log_{10} \left(\frac{R^2}{\text{MSE}}\right)
\end{equation}

where $R$ is the maximum possible pixel value (255 for 8-bit images).

\textbf{Structural Similarity Index (SSIM)} evaluates perceived changes in structural information, luminance, and contrast \cite{wang_image_2004}:

\begin{equation}
\text{SSIM}(x,y) = \frac{(2\mu_x \mu_y + a)(2\sigma_{xy} + b)}{(\mu_x^2 + \mu_y^2 + a)(\sigma_x^2 + \sigma_y^2 + b)}
\end{equation}

where $\mu_x$, $\mu_y$ are average pixel values, $\sigma_x^2$, $\sigma_y^2$ are variances, $\sigma_{xy}$ is covariance, and $a$, $b$ are stabilization constants.

\section{Experimental Results and Discussion}\label{result}

Comprehensive experiments evaluated all combinations of three wavelet families, five decomposition levels (1-5), two threshold methods (Bayes and Universal), and two thresholding techniques (Hard and Soft) across three noise levels ($\sigma = 10, 15, 25$). PSNR and SSIM values were calculated for each denoised image relative to its clean ground truth, with results averaged across the entire dataset for statistical reliability.

The optimal configurations and corresponding performance metrics are summarized in Table \ref{tab:wavelet-summary}.

\begin{table}[htbp!]
\centering
\caption{Optimal Wavelet Configurations and Performance}
\label{tab:wavelet-summary}
\begin{tabular}{llcccccc}
\toprule
\textbf{Noise ($\sigma$)} & \textbf{Method} & 
\textbf{\begin{tabular}[c]{@{}c@{}}Opt.\\ Decomp.\\ Level\end{tabular}} &
\textbf{\begin{tabular}[c]{@{}c@{}}Opt.\\ Wavelet\end{tabular}} &
\textbf{\begin{tabular}[c]{@{}c@{}}Opt.\\ Thr.\end{tabular}} &
\textbf{PSNR (dB)} & \textbf{SSIM} \\
\midrule
\multirow{2}{*}{\textbf{10}} 
  & Bayes      & 2 & \texttt{db3}     & Hard & $25.644 \pm 1.999$ & $0.606 \pm 0.120$ \\
  & Universal  & 2 & \texttt{bior6.8} & Hard & \textbf{$27.382 \pm 1.922$} & \textbf{$0.647 \pm 0.118$} \\
\midrule
\multirow{2}{*}{\textbf{15}} 
  & Bayes      & 2 & \texttt{bior6.8} & Hard & $23.865 \pm 1.790$ & $0.556 \pm 0.114$ \\
  & Universal  & 2 & \texttt{bior6.8} & Hard & \textbf{$25.253 \pm 1.812$} & \textbf{$0.589 \pm 0.113$} \\
\midrule
\multirow{2}{*}{\textbf{25}} 
  & Bayes      & 3 & \texttt{bior6.8} & Hard & $21.796 \pm 1.763$ & $0.493 \pm 0.103$ \\
  & Universal  & 3 & \texttt{bior6.8} & Hard & \textbf{$22.837 \pm 1.796$} & \textbf{$0.518 \pm 0.104$} \\
\bottomrule
\end{tabular}
\end{table}

\subsection{Key Findings}

The comprehensive analysis reveals several important findings:

\textbf{Universal Method Superiority}: The Universal thresholding method consistently outperformed the Bayes method across all noise levels, with performance gaps ranging from approximately 1.0 dB at $\sigma=25$ to 1.7 dB at $\sigma=10$. This robust result demonstrates that the global, non-adaptive threshold of the Universal method provides superior noise-signal discrimination compared to the adaptive BayesShrink approach for MRI images.

\textbf{Optimal Wavelet Configuration}: The biorthogonal wavelet (bior6.8) with hard thresholding emerged as the dominant configuration in five out of six scenarios, particularly at higher noise levels ($\sigma=15$ and $\sigma=25$). The linear phase and symmetric properties of the biorthogonal wavelet prove particularly effective for preserving structural integrity in MRI scans.

\textbf{Decomposition Level Optimization}: Lower decomposition levels (2-3) consistently yielded optimal results, suggesting that deeper decompositions may discard valuable signal information along with noise.

\textbf{Performance Limitations}: While wavelet methods provide substantial improvement over noisy images, the degradation of SSIM to approximately 0.5 at $\sigma=25$ highlights the inherent limitations of traditional wavelet approaches, particularly in preserving structural similarity at high noise levels.

\begin{figure}[!htb]
    \centering
    \includegraphics[width=0.8\linewidth]{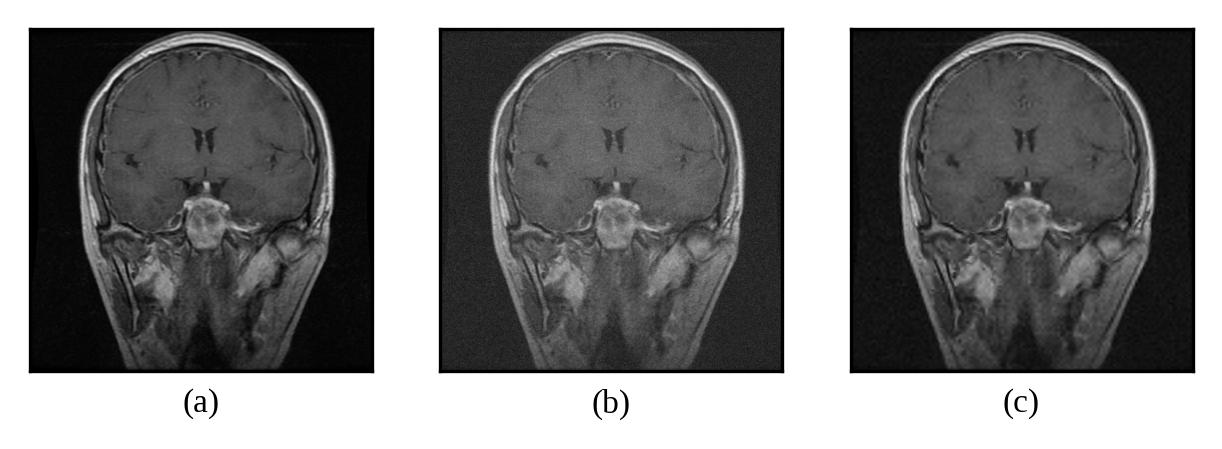}
    \caption{(a) Original Image, (b) Noisy Image($\mu = 0$, $\sigma = 10$), (c) Denoised Image (with threshold value $\tau_{bayes}$)}
    \label{fig:bayes_noise_sigma_10_sample1}
\end{figure}

\begin{figure}[!htb]
    \centering
    \includegraphics[width=0.8\linewidth]{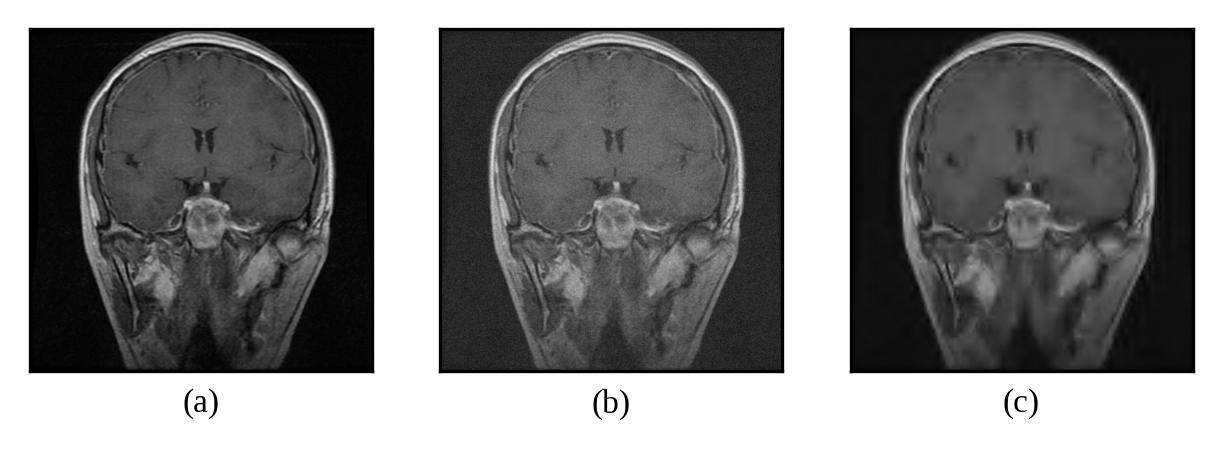}
    \caption{(a) Original Image, (b) Noisy Image($\mu = 0$, $\sigma = 10$), (c) Denoised Image (with threshold value $\tau_{universal}$)}
    \label{fig:universal_noise_sigma_10_sample1}
\end{figure}

\begin{figure}[!htb]
    \centering
    \includegraphics[width=0.8\linewidth]{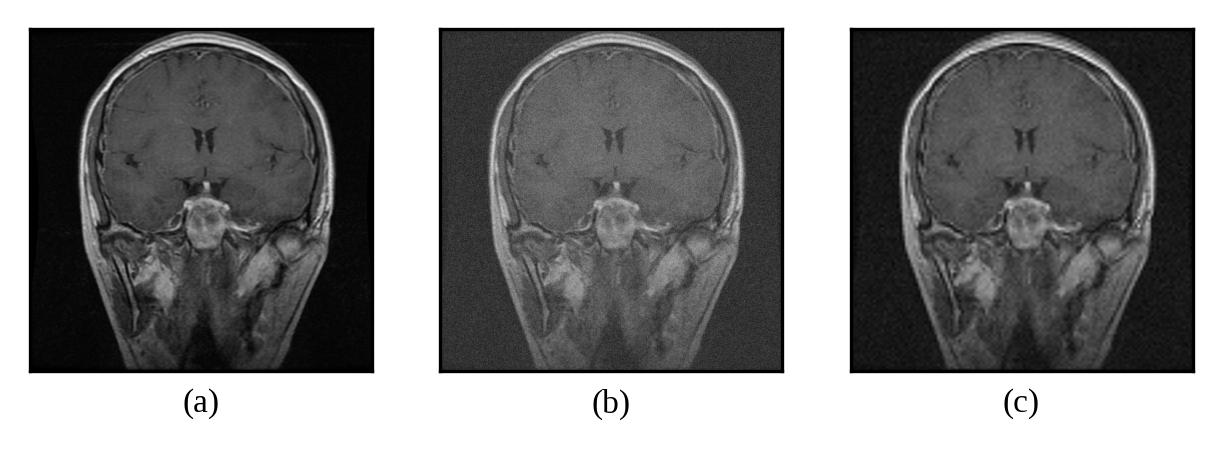}
    \caption{(a) Original Image, (b) Noisy Image($\mu = 0$, $\sigma = 15$), (c) Denoised Image (with threshold value $\tau_{bayes}$)}
    \label{fig:bayes_noise_sigma_15_sample1}
\end{figure}

\begin{figure}[!htb]
    \centering
    \includegraphics[width=0.8\linewidth]{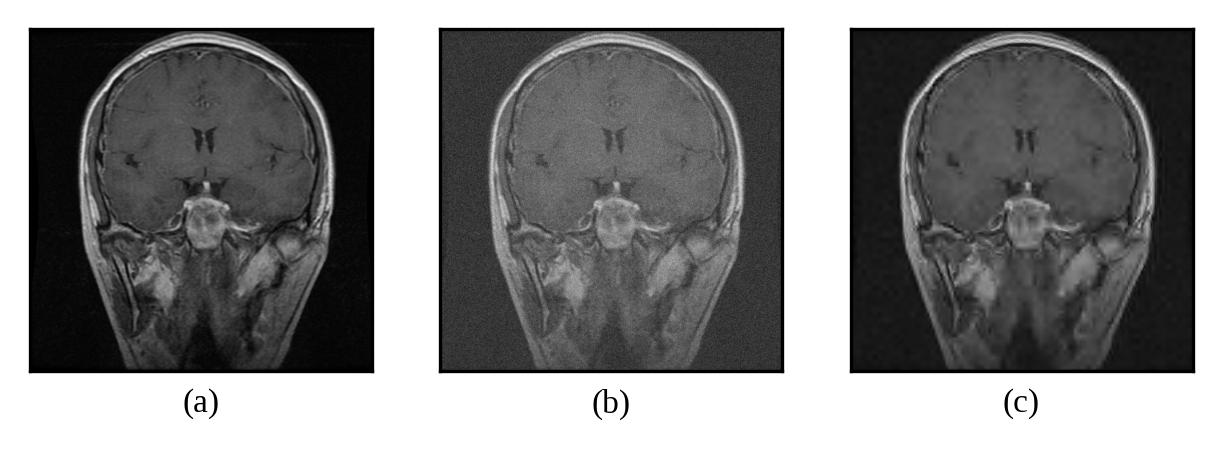}
    \caption{(a) Original Image, (b) Noisy Image($\mu = 0$, $\sigma = 15$), (c) Denoised Image (with threshold value $\tau_{universal}$)}
    \label{fig:universal_noise_sigma_15_sample1}
\end{figure}

\begin{figure}[!htb]
    \centering
    \includegraphics[width=0.8\linewidth]{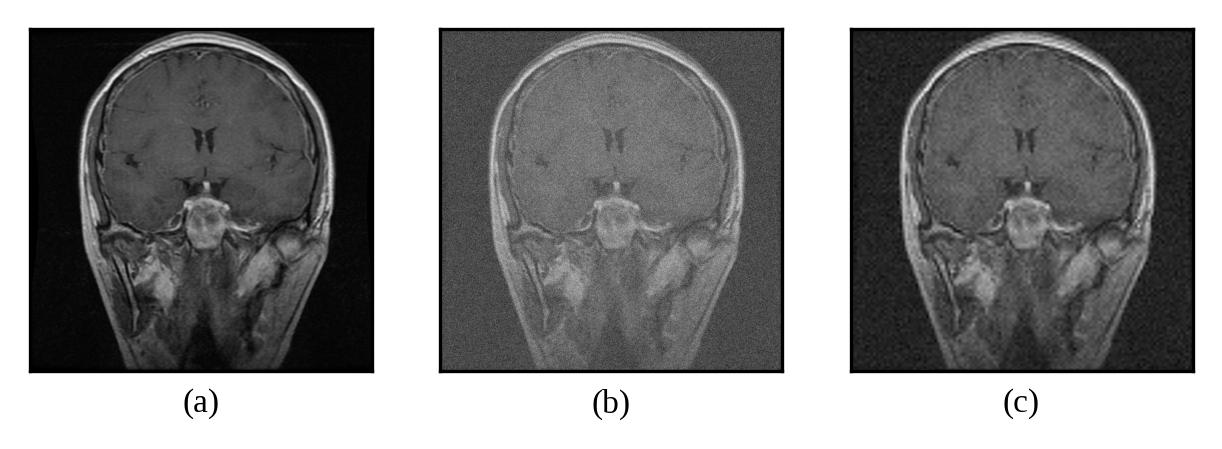}
    \caption{(a) Original Image, (b) Noisy Image($\mu = 0$, $\sigma = 25$), (c) Denoised Image (with threshold value $\tau_{bayes}$)}
    \label{fig:bayes_noise_sigma_25_sample1}
\end{figure}

\begin{figure}[!htb]
    \centering
    \includegraphics[width=0.8\linewidth]{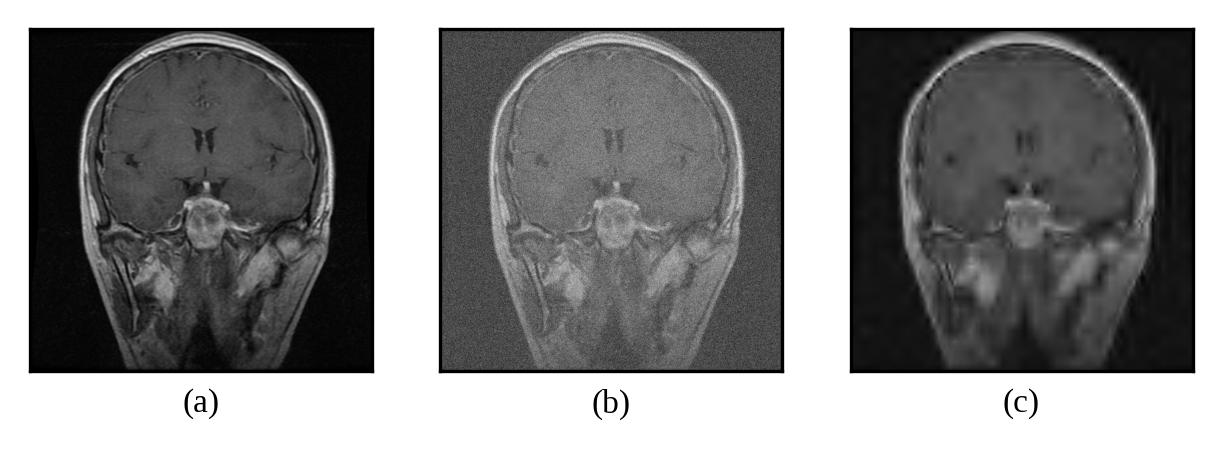}
    \caption{(a) Original Image, (b) Noisy Image($\mu = 0$, $\sigma = 25$), (c) Denoised Image (with threshold value $\tau_{universal}$)}
    \label{fig:universal_noise_sigma_25_sample1}
\end{figure}

Representative denoising results are illustrated in Figures \ref{fig:bayes_noise_sigma_10_sample1} through \ref{fig:universal_noise_sigma_25_sample1}, demonstrating the visual improvement achieved by both thresholding methods across different noise levels.

\section{Conclusion}\label{conclusion}

This work presented a systematic and comprehensive evaluation of wavelet-based denoising for brain MRI images. Through extensive experimentation across a large dataset, we established reliable performance benchmarks and identified optimal parameter configurations for various noise conditions. The Universal thresholding method with bior6.8 wavelet and hard thresholding represents the most effective configuration for this application, consistently outperforming alternative approaches.

While wavelet-based methods provide significant improvement over noisy images with computational efficiency and interpretability advantages, their non-adaptive nature fundamentally constrains performance, particularly at high noise levels. The quantitative benchmarks and insights established in this work provide essential groundwork for subsequent research phases, which will explore whether the data-driven, adaptive capabilities of deep learning approaches can overcome these limitations and achieve superior denoising performance in biomedical imaging applications.

\begin{thebibliography}{10}

\bibitem{kollem_novel_2023}
Sreedhar Kollem, Katta~Ramalinga Reddy, and Duggirala~Srinivasa Rao.
\newblock A novel diffusivity function-based image denoising for {MRI} medical images.
\newblock {\em Multimed Tools Appl}, 82(21):32057--32089, September 2023.

\bibitem{guo_review_2022}
Tiantian Guo, Tongpo Zhang, Enggee Lim, Miguel López-Benítez, Fei Ma, and Limin Yu.
\newblock A {Review} of {Wavelet} {Analysis} and {Its} {Applications}: {Challenges} and {Opportunities}.
\newblock {\em IEEE Access}, 10:58869--58903, 2022.

\bibitem{patil_efficient_2023}
Rajesh Patil and Surendra Bhosale.
\newblock Efficient {Denoising} of {Multi}-modal {Medical} {Image} using {Wavelet} {Transform} and {Singular} {Value} {Decomposition}.
\newblock In {\em 2023 {IEEE} {IAS} {Global} {Conference} on {Emerging} {Technologies} ({GlobConET})}, pages 1--6, London, United Kingdom, May 2023. IEEE.

\bibitem{zhang_study_2022}
Yanzhu Zhang, Tingting Liu, Fan Yang, and Qi~Yang.
\newblock A {Study} of {Adaptive} {Fractional}-{Order} {Total} {Variational} {Medical} {Image} {Denoising}.
\newblock {\em Fractal Fract}, 6(9):508, September 2022.

\bibitem{el-shafai_efficient_2022}
Walid El-Shafai, Samy Abd El-Nabi, El-Sayed M.~El-Rabaie, Anas M.~Ali, Naglaa F.~Soliman, Abeer D.~Algarni, and Fathi E.~Abd El-Samie.
\newblock Efficient {Deep}-{Learning}-{Based} {Autoencoder} {Denoising} {Approach} for {Medical} {Image} {Diagnosis}.
\newblock {\em Computers, Materials \& Continua}, 70(3):6107--6125, 2022.

\bibitem{izadi_image_2023}
Saeed Izadi, Darren Sutton, and Ghassan Hamarneh.
\newblock Image denoising in the deep learning era.
\newblock {\em Artif Intell Rev}, 56(7):5929--5974, July 2023.

\bibitem{tian_multi-stage_2023}
Chunwei Tian, Menghua Zheng, Wangmeng Zuo, Bob Zhang, Yanning Zhang, and David Zhang.
\newblock Multi-stage image denoising with the wavelet transform.
\newblock {\em Pattern Recognition}, 134:109050, February 2023.

\bibitem{gondara_medical_2016}
Lovedeep Gondara.
\newblock Medical image denoising using convolutional denoising autoencoders.
\newblock In {\em 2016 {IEEE} 16th {International} {Conference} on {Data} {Mining} {Workshops} ({ICDMW})}, pages 241--246, December 2016.
\newblock arXiv:1608.04667 [cs].

\bibitem{jun_cheng_brain_2017}
{Jun Cheng}.
\newblock Brain {Tumor} {Dataset}, April 2017.

\bibitem{chipman_adaptive_1997}
Hugh~A. Chipman, Hugh~A. Chipman, Eric~D. Kolaczyk, and Robert~E. McCulloch.
\newblock Adaptive bayesian wavelet shrinkage.
\newblock {\em Journal of the American Statistical Association}, 92(440):1413--1421, December 1997.

\bibitem{donoho_ideal_1994}
David~L Donoho and Iain~M Johnstone.
\newblock Ideal spatial adaptation by wavelet shrinkage.
\newblock {\em Biometrika}, 81(3):425--455, September 1994.

\bibitem{donoho_-noising_1995}
D.~L. Donoho.
\newblock De-noising by soft-thresholding.
\newblock {\em IEEE Trans. Inf. Theor.}, 41(3):613--627, May 1995.

\bibitem{wang_mean_2009}
Zhou Wang and Alan~C. Bovik.
\newblock Mean squared error: {Love} it or leave it? {A} new look at {Signal} {Fidelity} {Measures}.
\newblock {\em IEEE Signal Processing Magazine}, 26(1):98--117, January 2009.

\bibitem{wang_image_2004}
Zhou Wang, A.C. Bovik, H.R. Sheikh, and E.P. Simoncelli.
\newblock Image quality assessment: from error visibility to structural similarity.
\newblock {\em IEEE Transactions on Image Processing}, 13(4):600--612, April 2004.

\bibitem{majeed_zangana_classical_2024}
Hewa Majeed~Zangana and Firas~Mahmood Mustafa.
\newblock From {Classical} to {Deep} {Learning}: {A} {Systematic} {Review} of {Image} {Denoising} {Techniques}.
\newblock {\em JICS}, 3(1):50--65, July 2024.

\bibitem{khader_denoising_2023}
Firas Khader, Gustav Müller-Franzes, Soroosh Tayebi~Arasteh, Tianyu Han, Christoph Haarburger, Maximilian Schulze-Hagen, Philipp Schad, Sandy Engelhardt, Bettina Baeßler, Sebastian Foersch, Johannes Stegmaier, Christiane Kuhl, Sven Nebelung, Jakob~Nikolas Kather, and Daniel Truhn.
\newblock Denoising diffusion probabilistic models for {3D} medical image generation.
\newblock {\em Sci Rep}, 13(1):7303, May 2023.

\bibitem{sahu_application_2023}
Alpana Sahu, K.~P.~S. Rana, and Vineet Kumar.
\newblock An application of deep dual convolutional neural network for enhanced medical image denoising.
\newblock {\em Med Biol Eng Comput}, 61(5):991--1004, May 2023.

\bibitem{shukla_effective_2023}
Abhay Shukla, K.~Seethalakshmi, P~Hema, and Jitendra~Chandrakant Musale.
\newblock An {Effective} {Approach} for {Image} {Denoising} {Using} {Wavelet} {Transform} {Involving} {Deep} {Learning} {Techniques}.
\newblock In {\em 2023 4th {International} {Conference} on {Smart} {Electronics} and {Communication} ({ICOSEC})}, pages 1381--1386, Trichy, India, September 2023. IEEE.

\bibitem{atal_optimal_2023}
Dinesh~Kumar Atal.
\newblock Optimal {Deep} {CNN}–{Based} {Vectorial} {Variation} {Filter} for {Medical} {Image} {Denoising}.
\newblock {\em J Digit Imaging}, 36(3):1216--1236, January 2023.

\bibitem{muller-franzes_multimodal_2023}
Gustav Müller-Franzes, Jan~Moritz Niehues, Firas Khader, Soroosh~Tayebi Arasteh, Christoph Haarburger, Christiane Kuhl, Tianci Wang, Tianyu Han, Teresa Nolte, Sven Nebelung, Jakob~Nikolas Kather, and Daniel Truhn.
\newblock A multimodal comparison of latent denoising diffusion probabilistic models and generative adversarial networks for medical image synthesis.
\newblock {\em Sci Rep}, 13(1):12098, July 2023.

\bibitem{kaur_complete_2023}
Amandeep Kaur and Guanfang Dong.
\newblock A {Complete} {Review} on {Image} {Denoising} {Techniques} for {Medical} {Images}.
\newblock {\em Neural Process Lett}, 55(6):7807--7850, December 2023.

\bibitem{ma_strunet_2023}
Yuhui Ma, Qifeng Yan, Yonghuai Liu, Jiang Liu, Jiong Zhang, and Yitian Zhao.
\newblock {StruNet}: {Perceptual} and low‐rank regularized transformer for medical image denoising.
\newblock {\em Medical Physics}, 50(12):7654--7669, December 2023.

\bibitem{gautam_novel_2023}
Divya Gautam, Kavita Khare, and Bhavana~P. Shrivastava.
\newblock A {Novel} {Guided} {Box} {Filter} {Based} on {Hybrid} {Optimization} for {Medical} {Image} {Denoising}.
\newblock {\em Applied Sciences}, 13(12):7032, June 2023.

\bibitem{patil_medical_2022}
Rajesh Patil and Surendra Bhosale.
\newblock Medical {Image} {Denoising} {Techniques}: {A} {Review}.
\newblock {\em IJONEST}, 4(1):21--33, January 2022.

\bibitem{kollem_image_2022}
Sreedhar Kollem, Katta Ramalinga~Reddy, Duggirala Srinivasa~Rao, Chintha Rajendra~Prasad, V.~Malathy, J.~Ajayan, and Deboraj Muchahary.
\newblock Image denoising for magnetic resonance imaging medical images using improved generalized cross‐validation based on the diffusivity function.
\newblock {\em Int J Imaging Syst Tech}, 32(4):1263--1285, July 2022.

\bibitem{mohd_sagheer_review_2020}
Sameera~V. Mohd~Sagheer and Sudhish~N. George.
\newblock A review on medical image denoising algorithms.
\newblock {\em Biomedical Signal Processing and Control}, 61:102036, August 2020.

\bibitem{jifara_medical_2019}
Worku Jifara, Feng Jiang, Seungmin Rho, Maowei Cheng, and Shaohui Liu.
\newblock Medical image denoising using convolutional neural network: a residual learning approach.
\newblock {\em J Supercomput}, 75(2):704--718, February 2019.

\bibitem{sarah_survey_2025}
Benziane Sarâh.
\newblock Survey – {Start} with {Image} {Denoising}.
\newblock {\em WSEAS TRANSACTIONS ON SIGNAL PROCESSING}, 21:41--50, April 2025.

\bibitem{gor_two_2024}
Ashishkumar Gor and C.K. Bhensdadia.
\newblock Two self-supervised image denoiser designs with discrete wavelet transform and non-local means-based algorithms.
\newblock {\em 2576-8484}, 8(6), December 2024.

\bibitem{zangana_review_2024}
Hewa~Majeed Zangana and Firas~Mahmood Mustafa.
\newblock Review of {Hybrid} {Denoising} {Approaches} in {Face} {Recognition}: {Bridging} {Wavelet} {Transform} and {Deep} {Learning}.
\newblock {\em ijcs}, 13(4), July 2024.

\bibitem{kascenas_role_2023}
Antanas Kascenas, Pedro Sanchez, Patrick Schrempf, Chaoyang Wang, William Clackett, Shadia~S. Mikhael, Jeremy~P. Voisey, Keith Goatman, Alexander Weir, Nicolas Pugeault, Sotirios~A. Tsaftaris, and Alison~Q. O’Neil.
\newblock The role of noise in denoising models for anomaly detection in medical images.
\newblock {\em Medical Image Analysis}, 90:102963, December 2023.

\bibitem{yuan_hybrid_2024}
Qikun Yuan.
\newblock Hybrid {Machine} {Learning} {Techniques} for {Image} {Denoising} {Based} on {Wavelet} {Transform}.
\newblock In {\em 2024 {IEEE} 6th {International} {Conference} on {Power}, {Intelligent} {Computing} and {Systems} ({ICPICS})}, pages 1162--1169, Shenyang, China, July 2024. IEEE.

\bibitem{liu_multi-level_2018}
Pengju Liu, Hongzhi Zhang, Kai Zhang, Liang Lin, and Wangmeng Zuo.
\newblock Multi-level {Wavelet}-{CNN} for {Image} {Restoration}, 2018.
\newblock Version Number: 2.

\bibitem{xu_wavelet_2024}
Ruotao Xu, Yong Xu, Xuhui Yang, Haoran Huang, Zhenghua Lei, and Yuhui Quan.
\newblock Wavelet analysis model inspired convolutional neural networks for image denoising.
\newblock {\em Applied Mathematical Modelling}, 125:798--811, January 2024.

\bibitem{zangana_hybrid_2024}
Hewa~Majeed Zangana and Firas~Mahmood Mustafa.
\newblock Hybrid {Image} {Denoising} {Using} {Wavelet} {Transform} and {Deep} {Learning}.
\newblock {\em EAI Endorsed Trans AI Robotics}, 3, November 2024.

\bibitem{liu_densely_2020}
Wei Liu, Qiong Yan, and Yuzhi Zhao.
\newblock Densely {Self}-guided {Wavelet} {Network} for {Image} {Denoising}.
\newblock In {\em 2020 {IEEE}/{CVF} {Conference} on {Computer} {Vision} and {Pattern} {Recognition} {Workshops} ({CVPRW})}, pages 1742--1750, Seattle, WA, USA, June 2020. IEEE.

\bibitem{tian_deep_2020}
Chunwei Tian, Lunke Fei, Wenxian Zheng, Yong Xu, Wangmeng Zuo, and Chia-Wen Lin.
\newblock Deep learning on image denoising: {An} overview.
\newblock {\em Neural Networks}, 131:251--275, November 2020.

\bibitem{benhassine_medical_2021}
Nasser~Edinne Benhassine, Abdelnour Boukaache, and Djalil Boudjehem.
\newblock Medical image denoising using optimal thresholding of wavelet coefficients with selection of the best decomposition level and mother wavelet.
\newblock {\em International Journal of Imaging Systems and Technology}, 31(4):1906--1920, 2021.
\newblock \_eprint: https://onlinelibrary.wiley.com/doi/pdf/10.1002/ima.22589.

\bibitem{lehtinen_noise2noise_2018}
Jaakko Lehtinen, Jacob Munkberg, Jon Hasselgren, Samuli Laine, Tero Karras, Miika Aittala, and Timo Aila.
\newblock {Noise2Noise}: {Learning} {Image} {Restoration} without {Clean} {Data}, October 2018.
\newblock arXiv:1803.04189 [cs].

\bibitem{zhang_plug-and-play_2021}
Kai Zhang, Yawei Li, Wangmeng Zuo, Lei Zhang, Luc~Van Gool, and Radu Timofte.
\newblock Plug-and-{Play} {Image} {Restoration} with {Deep} {Denoiser} {Prior}, July 2021.
\newblock arXiv:2008.13751 [eess].

\bibitem{liang_swinir_2021}
Jingyun Liang, Jiezhang Cao, Guolei Sun, Kai Zhang, Luc~Van Gool, and Radu Timofte.
\newblock {SwinIR}: {Image} {Restoration} {Using} {Swin} {Transformer}, August 2021.
\newblock arXiv:2108.10257 [eess].

\bibitem{bin_rahman_mitigating_2025}
Asadullah Bin~Rahman, Masud Ibn~Afjal, and Md.~Abdulla Al~Mamun.
\newblock Mitigating {Noise} from {Biomedical} {Images} {Using} {Wavelet} {Transform} {Techniques}.
\newblock In {\em 2025 {International} {Conference} on {Electrical}, {Computer} and {Communication} {Engineering} ({ECCE})}, pages 1--6, February 2025.

\end{thebibliography}

\end{document}